

\font\tafont =  cmbx10
\font\tbfont =  cmbx9

\magnification =1200

  \def\titlea#1#2{{~ \vskip 2 truecm
  \tafont {\centerline  {#1}} 
  {\centerline{#2}}   }
 \vskip 3truecm
 }

\def \pn{\par\noindent}
\def \bs{\bigskip}
\def\titleb#1{\pn \bigskip \bigskip\parindent=0pt {\tbfont {#1}  } \bigskip
\parindent=16pt}
\def\Ref{\bs\pn{{\bf References}}\smallskip \parskip=5 pt\parindent=0pt}

\hfuzz=2pt
\tolerance=500
\abovedisplayskip=2 mm plus6pt minus 4pt
\belowdisplayskip=2 mm plus6pt minus 4pt
\abovedisplayshortskip=0mm plus6pt minus 2pt
\belowdisplayshortskip=2 mm plus4pt minus 4pt
\predisplaypenalty=0
\clubpenalty=10000
\widowpenalty=10000

\def \phi{\varphi}

\def \om{\omega}

\def \A{{\cal A}}
\def \B{{\cal B}}

\def \D{{\cal D}}
\def \a{\alpha}

\def\grad{\nabla}
\def\Ker{{\rm Ker}}

\def\({\left(}
\def\){\right)}

\def \d{{\rm d}}
\def\=#1{\bar #1}
\def\~#1{\widetilde #1}
\def\.#1{\dot #1}
\def\^#1{\widehat #1}
\def\"#1{\ddot #1}
\baselineskip 0.58 cm

\def \tr{transformation}
\def \com {constant of motion}
\def \coms {constants of motion}

{ \nopagenumbers

\titlea{On the convergence of normalizing \tr s}
{in the presence of symmetries}

\centerline{Giampaolo Cicogna}
\bigskip
\centerline{\it Dipartimento di Fisica, Universit\`a di Pisa}
\centerline{\it P.za Torricelli 2, I-56126 Pisa (Italy)}
\centerline{{\tt E-Mail: cicogna@ipifidpt.difi.unipi.it}}

\vskip 4cm
\pn
{\bf Abstract}.
\pn
It is shown that, under suitable conditions, involving in
particular the existence of analytic \coms , the presence of Lie point
symmetries can ensure the convergence of the transformation taking a
vector field (or dynamical system) into normal form.

\pn
{\tt A short version of this paper has been submitted, as a Letter, to
the Journal of Phys. A}.

\vfill\eject}

\titleb {1. Introduction.}

The technique of transforming a system of first-order ordinary differential
equations (also called  dynamical system) into
"normal form" is an old and well known method of investigation, going
back to the classical work of Poincar\'e (and then of Dulac and Birkhoff),
and developed in more recent times by several authors (see e.g. [2,3,5-7,22]
and ref. therein). Its connection with symmetry properties
\big(precisely, Lie-point symmetries (see [20,21])\big) of the dynamical
system have been also pointed out [1,6,11,12,16].
One of the main troubles with this procedure is given by the problem  of
the convergence of the normalizing \tr : it is known in fact that these
\tr s are performed by means of recursive techniques, and only
special conditions can ensure their convergence~~[7].

A possible way-out which is usually adopted is that of considering
these \tr s "up to some
finite order", i.e. of considering "approximate normal forms" (and,
correspondingly, approximate solutions, approximate symmetries, etc. of
the given system; see e.g. [4,14]). In this paper, instead, we are
dealing with the case
of converging normalizing \tr s, and with some conditions ensuring
convergence. Our discussion is  related to a theorem by
Markhashov [19]: even if the original proof of this theorem has appeared to
be not complete, a similar result has been proved at least in some
particular case (see [8]). A very remarkable result in this context is given
in a recent paper by Bruno and Walcher [10], in the case of 2-dimensional
systems. The main idea is that the presence of some symmetry
of the dynamical system can ensure - under suitable
conditions - the convergence of the normalizing \tr s.

Our paper presents some considerations in the same direction. In
particular, we give a direct proof of a "Markhashov-type" theorem in a well
definite and simple case, and analyze some applications of this idea, which
include a generalization of the Bruno and Walcher result~~[10].

\vfill\eject

\titleb{2. A "Markhashov-type" theorem.}

We will consider dynamical systems (DS) of the form
$$\dot u=f(u)\qquad \qquad u=u(t)\in R^n\eqno(1)$$
where $\dot u=\d u/\d t,\ f$ is assumed to be analytic in a neighbourhood
of $u=0$, with $f(0)=0$, and
$$f(u)=Au+F(u)\eqno(2)$$
where the matrix $A\equiv(\nabla f)(0)$ is assumed to be nonzero and
diagonalizable.  Let us remark that most of the results below
could be extended to the  non-diagonalizable case, apart from some
complications in the notations and statements (see  [1,13]).

As well known [2,7], a normalizing \tr\ is a nonlinear formal \tr :
$$u\to \^u\ =\ u\ +\ldots\eqno(3)$$
transforming (1-2) into a new DS which we write in the form
(to avoid cumbersome notations, we will denote by $u$ both the "original"
and the transformed coordinates)
$$\dot u = \^f(u)=Au+\^F(u)\eqno(4)$$
where the nonlinear part $\^F(u)$ is in "normal form" (NF). To define this
notion, one introduces in the space of analytic functions, defined in a
neighbourhood of $u=0$, the Lie-Poisson bracket
$$\{f,g\}_k=(f\cdot\grad)g_k-(g\cdot\grad)f_k \qquad \qquad (k=1,\ldots,n)
            \eqno(5)$$
and, given any $n\times n$ matrix $A$, the "homological operator" $\A$
$$\A(f)=\{Au,f\}=(Au)\cdot\grad f-Af \ .\eqno(6)$$
Then, a nonlinear vector function $\^F$ is said to be in NF with respect
to $A$ (or resonant with $A$) if
$$\A(\^F)=0 \eqno(7)$$
In the basis where $A$ is diagonal,
with eigenvalues $a_1,\ldots,a_n$, a monomial
$\^F_k(u)=u_1^{m_1}~\cdot~\ldots~\cdot~u_n^{m_n}$ of degree $j$ (with $m_i$
integer numbers such that $\sum_i m_i=j, \ m_i\ge 0$)
is resonant if $(m,a)\equiv\sum_i m_i a_i= a_k$. As well
known, the relevance of the above definitions is essentially due to the
fact that, given a vector function $f$, all nonresonant terms can be removed
by means of a formal coordinate transformation.

We also say that a vector function
$$g(u)=Bu+G(u)\eqno(8)$$
is a (Lie-point time-independent) symmetry for the DS (1-2) if
$$\{f,g\}=0\ .\eqno(9)$$
In terms of Lie algebras, one says that the vector field
operator $g\cdot\grad$  generates a symmetry of the DS.

A scalar function $\rho=\rho(u)$ is a (time-independent)
constant of motion (or first integral) for the DS (1-2) if
the Lie derivative along $f$ vanishes:
$${\d\rho\over{\d t}}\Big|_f\equiv f\cdot \grad \rho=0 \ .\eqno(10)$$
The above definitions of symmetry and of \com\ can be clearly applied
both to analytic functions and to formal power series.

Our discussion needs  few preliminary results, some of which are
rather simple or well known; however, for clarity and completeness, we give
all of them: some of these
introductory results may also have an independent interest.
\bs\pn
{\bf Lemma 1}. If $g$ is a symmetry for the DS, and $\rho$ a \com\
for it, then also
$$h=\rho\ g$$
is a symmetry for the DS. More precisely, the algebra of the symmetries
$g$ of a DS is {\it module} over the \coms\ of the DS.
\bs
The first part of this Lemma is immediate; the other statement describes
the general property [13,15,16,21] of the solutions $g$ to the system of
PDE's (9) which gives the symmetries of (1-2). See later on for some remark
concerning the number of the "admissible" \coms . A fundamental step in
the discussion is provided by the following Lemma.
\bs\pn
{\bf Lemma 2} [1,12,23]. If the DS (1-2) admits a symmetry $g$ (8), and $f$
is put in NF
$$f\to \^f=Au+\^F,\qquad \^F\in \Ker\A\eqno(11)$$
by a formal normalizing \tr , then $g$ is transformed into a new
form (not necessarily normal and possibly formal) which we denote by
(we reserve the notation $\,\^\cdot\, $ only to NF)
$$g\to \~g=Bu+\~G\eqno(12)$$
satisfying \big(together with $\{\^f,\~g\}=0$, following directly from
(9)\big)
$$\~G\in\Ker\A\qquad {\rm or\ equivalently}\qquad \{Au,\~G\}=0\eqno(13)$$
$$\^F\in\Ker\B\qquad {\rm or\ equivalently}\qquad \{Bu,\^F\}=0\eqno(14)$$
where $\B$ is the homological operator $\B(\cdot)=\{Bu,\cdot\}$.
\pn
{\bf Proof} (a sketch). Expanding $\^F$ in formal power series:
$$\^F=\sum_{j\ge 2}\^F_j$$
where $\^F_j$ are homogeneous polynomials of degree $j$, and expanding
in a similar way $\~G$, one has from $\{\^f,\~g\}=0$ at order 1
$$[A,B]=0\eqno(15)$$
and at order 2
$$\{Au,\~G_2\}=\{Bu,\^F_2\}\eqno(16)$$
Applying to this equality the operator $\A$, one has $\A^2(\~G_2)=0$
which gives, thanks to the assumption on the matrix $A$ (see also [12]),
$\A(\~G_2)=0$, and from (16), also $\B(\^F_2)=0$. Iterating the procedure,
one obtains (13-14).
\bs\pn
{\bf Remark} [12]. By means of a further formal \tr , one may also normalize
$\~g\to\^g$, thus obtaining the "joint" NF:
$$\^F,\^G\in \Ker\A\cap\Ker\B \ .\eqno(17)$$
\bs
An immediate but very important consequence of the notion of NF and
of Lemma 2 (in particular eq. (14)) is the following proposition.
\bs\pn
{\bf Proposition 1}. If the DS is in NF (11), then it admits the linear
symmetry $Au$. If the original DS (1-2) admits a symmetry $g=Bu+G$, then
the normalized DS (11) also admits the linear symmetry $Bu$ .
\bs
Let us now recall the two basic conditions which ensure that
a vector function $f=Au+F$ can be put in NF by a converging \tr\ [7], namely:
\bs\pn
{\sl Condition "A"}: there is a coordinate \tr\ changing $f$ to $\^f$,
where $\^f$ has the form
$$\^f=Au+\a(u)Au$$
and $\a(u)$ is some scalar-valued power series  (with $\a(0)=0$).
\bs\pn
{\sl Condition "$\om$"}: let $\om_k=\min|(q,a)|$ for all positive
integers $q_i$ such that $\sum_{i=1}^n q_i<2^k$ and $(q,a)\ne 0$: then the
series
$$\sum_{k=1}^\infty 2^{-k}\ln \om_k$$
is convergent.
\bs\pn
The first condition will play a key role in our discussion; the other one
is a weaker condition, controlling the
appearance of small divisors [7], and we explicitly assume that it is
always satisfied here. In particular, it is satisfied in the
cases considered in sect. 4.

We are now in position to give a simple and direct proof of the following
result. Let us preliminarily
note that any DS admits an obvious symmetry, namely $g\equiv f$,
which is in fact the generator of the dynamical
flow. Accordingly, when considering the symmetries of a DS, it is always
understood that none of them (nor their linear combinations) is
proportional to $f$.
Also, we may clearly exclude the case (which may be considered here as
"trivial") that the DS, once in
NF, takes the form $\.u=\big(1+\a(u)\big)Au$: in this case indeed the
Condition "A" is satisfied, and the convergence of the normalizing \tr\
automatically guaranteed.
\bs\pn
{\bf Theorem 1}. Assume that the DS (1-2) possesses a finite number
$\ell$ ($\ge 1$) of analytic symmetries $g_j=B_ju+G_j$, where all
the matrices $B_j$  are linearly independent (and not zero). Assume also
that, once in NF, the DS admits exactly $\ell$ linearly independent
(possibly formal) symmetries. Then, there exists a converging normalizing
\tr\ for $f$.
\pn
{\bf Proof}. Let us start by writing the DS into NF by means of some formal
normalizing \tr
$$f\to\^f=Au+\^F$$
Under this \tr\ the $\ell$ symmetries become
$$g_j\to \~g_j=B_ju+\~G_j\eqno(18)$$
According to Prop. 1, the linear parts $B_ju$ together with $Au$ are
$(\ell +1)$ symmetries of the DS in NF; excluding the trivial case
$\^f=Au$, this implies that there must be a linear combination
of the $B_j$ which is equal to $A$: it is not restrictive to assume
e.g. $B_1=A$; consider then
$$\~g_1=Au+\~G_1$$
By assumption, our problem in NF admits exactly $\ell$ symmetries (plus the
trivial one $\^f$), therefore $\~g_1$ must be a linear combination
(with constant coefficients) of the symmetries at disposal, i.e.
of $s_j=B_ju$ and $\^f$. But the linear part of $\~g_1$ is just $Au$,
and this forces $\~G_1=0$, or
$$\~g_1=Au$$
This implies that Condition "A" is satisfied by the \tr\ of the symmetry
$g_1=Au+G_1\to\~g_1=Au$; then
there is a normalizing \tr\ which is convergent. Under this \tr ,
$f$ is transformed into $\^f=Au+\^F$ which is in NF, according to Lemma 2
(indeed $\^F\in\Ker\A$, being $g_1=Au+G_1$).

\bs
The above theorem looks quite "formal"  and not easily applicable in
concrete cases: in fact, it may be difficult to check in practice that
the required properties of the NF (which is usually  "a priori"
not explicitly known)
are verified.  In the two next sections, we will give more concrete
versions of this result
and  study some cases in which the above hypotheses can be fulfilled.

\titleb{3. The number of \coms\ and symmetries of a DS.}

First of all, let us remark that one of the crucial hypotheses of
Theorem 1 is that the DS in NF admits a finite number of independent
symmetries. According to Lemma 1, the finiteness of this number depends
in an essential way on the number of independent \coms\ of the DS.
In fact, it is clear that the presence of \coms\ of the problem in NF
precludes the application of our above argument: indeed,
the nonlinear part $\~G_1$ of the symmetry $\~g_1$ could  be obtained
as a combination of $\^f$ and of the other $B_ju$ multiplied by
suitable \coms , and Condition "A" for the symmetry $g_1$
fails to be verified.

To carefully discuss this point, let us recall the two following
relevant results.
\bs\pn
{\bf Lemma 3} [10,23]. If $\^f=Au+\^F$ is in NF, i.e. $\^F\in \Ker \A$, then
any (formal) \com\ $\rho$ of the DS $\dot u=\^f$ is also a (formal)
\com\ of the linear part $\dot u=Au$.
\bs\pn
{\bf Lemma 4} [13,16]. Given the matrix $A$, the set $\Ker \A$ of
terms $\^F(u)$ resonant  with $A$, is given by $M(\mu(u))u$, where $M$ is
the most general matrix such that $[M,A]=0$ and its entries $M_{ij}$
are functions of the  constants
of motion $\mu=\mu(u)$ of the linear system $\.u=Au$; or also, choosing a
basis $M_j$ in the space of the matrices $M$ commuting with $A$,
the most general $\^F$ in NF is
$$\^F=\sum_j\mu_j(u)M_ju \ . \eqno(19)$$
with $\mu_j(0)=0$.
\bs
According to Lemma 3, there are two essentially different ways in which
a DS may admit a finite number of \coms , namely:
\pn
1) the linear part $\dot u=Au$ admits a finite number of linearly
independent \coms\ (then the same is true, {\it a fortiori}, for the full DS);
\pn
2) the linear part does admit infinite \coms\ (functionally dependent, of
course), but only a finite number of them (possibly none) is admitted by
the full DS.

Both these cases deserve some remarks.

Consider, as an example of case 1), a 2-dimensional system with $A={\rm
diag}(1,2)$: then a \com\ is  $\rho=u_1^2/u_2$. This is
{\sl not} an analytic nor formal  \com , but it can be admitted in this
context, because we are interested in analytic symmetries, which are
described by vector functions of the form $g=Bu+G$, i.e. with $g(0)=0$: then
$\rho g$ may be analytic even if $\rho$ is not.  In the
hypotheses of Theorem 1 we assumed $B_j\ne 0$, then the only admitted \coms\
are of the type $u_1^{m_1}\cdot\ldots\cdot u_n^{m_n}$ where  {\sl one at most}
of the $m_i$ may be $-1$. Therefore, only a finite number of independent
\coms\ are admitted in this case. A situation of this type
happens e.g. when the resonant eigenvalues are real and have the same sign.
{}From the point of view of the convergence of the normalizing \tr s, this
case is actually more conveniently treated with the notion of the
Poincar\'e domain (which provides direct criteria of convergence [2]); the
present discussion offers simply a different approach to the problem.

A more interesting situation occurs in case 2), which we are going to
consider.

Assume e.g. that the DS takes in NF the following special form (where
$-$ in the sum appearing in (19) $-$ only the  matrix $A$
and some other matrix $M$, commuting with $A$ and not proportional to $A$,
are present):
$$\dot u=\^f=Au+\a(u)Au+\mu(u)Mu\eqno(20)$$
According to Lemma 3, any \com\ of (20) is also
\com\ of the linear problem $\dot u=Au$: therefore, to find a \com\ of
(20), we start with a \com\ $\rho$ of the linear problem,
i.e. $Au\cdot \grad \rho=0$, then one gets
$${\d\rho\over{\d t}}\Big|_{\^f}=\^f\cdot\grad \rho=
              \mu(u)Mu\cdot\grad \rho \eqno(21)$$
Now, $Mu\cdot\grad\rho=0$ is verified if and only if $\rho$ is a \com\
of the linear problem
$\dot u=Mu$; therefore, if there are no common \coms\ of the problems
$\dot u=Au$ and $\dot u=Mu$, one has $\d\rho/\d t|_{\^f}\ne 0$.
In particular, this happens
if $M=I\ ( =$ Identity) and the eigenvalues of $A$ are nondegenerate:
indeed the
\coms\ of $\.u=Iu$ are of the form $u_i/u_j$ (or
functions thereof), but the assumption on the eigenvalues of $A$ excludes
that these can be \coms\ of $\dot u=Au$. Then  we can state the following:
\bs\pn
{\bf Theorem $1'$}. Assume that: $i$) the DS (1-2) possesses a finite number
$\ell$ ($\ge 1$) of analytic symmetries $g_j=B_ju+G_j$, where all
the matrices $B_j$  are linearly independent (and not zero),
and where $\ell$ is precisely the number of the linearly
independent linear symmetries admitted by the DS once in NF;
\pn
$ii$) once in NF, the DS has the form (20), and the two linear
problems $\dot u=Au$ and $\dot u=Mu$ do not admit common \coms .
\pn
Then, the DS can be put in NF by means of a convergent normalizing \tr .
\bs\pn
{\bf Theorem $1''$}. Condition $i$) in Theorem $1'$ can be substituted by one
of the two following:
\pn
$i_1$) the DS (1-2) admits a nontrivial analytic symmetry $g=Bu+G$ such that
$B\ne 0$ is proportional to $A$;
\pn
$i_2$) the DS (1-2) admits an analytic symmetry $g=G(u)$ with vanishing
linear part and $G$ not proportional to $F$.
\bs\pn
{\bf Remarks}. a) Assumption $ii$) in Theorem $1'$ ensures that the NF (20)
has no other symmetry apart from the linear ones $B_ju$. Instead,
in the "trivial" case that Condition "A" is satisfied by
the DS, so that the NF has the form $\.u=\big(1+\a(u)\big)Au$,  any
\com\ of the linear problem $\dot u=Au$ is clearly also \com\ of the full
problem in NF. Then, in this case $-$ in contrast with the case covered by
Theorem $1'$ $-$ the DS in NF does admit, in general,
an infinite number of (functionally dependent) polynomial \coms , and
of symmetries as well.
\pn
b) In the assumption $ii$) of Theorem $1'$ the request that no common \coms\
are present can be substituted by the request that the only common
\coms\ are rational functions of degree $0$. Our final example is an
application of this assumption.
\pn
c) In the case the DS satisfies condition $i_2$) of Theorem $1''$, it is
clearly sufficient to consider the new nontrivial symmetry $g=G+f=Au+F+G$
to recover $i_1$). The argument then proceeds as in Theorem 1.

\titleb{4. Applications and examples.}

We want first to show that the theorem by Bruno and Walcher [10]
for 2-dimensional DS can be reobtained as a corollary of the above approach.
We have in fact:
\bs\pn
{\bf Corollary}. Consider a 2-dimensional DS $\.u=Au+F$ such that the
eigenvalues
$a_1,a_2$ of the matrix $A$ satisfy a relation $k_1a_1+k_2a_2=0$ where
$k_1,k_2$ are non-negative relatively prime integers, not simultaneously
zero. Then, if the DS possesses an analytic symmetry, it can be put in
NF by means of a converging \tr .
\pn
{\bf Proof}. Let $\.u=Au+F$ the DS, and $g=Bu+G$ its symmetry. If
$B=0$ then consider the new symmetry $g+f=Au+(F+G)$, so that the linear
part is now $\ne 0$. From $[A,M]=0$ and observing that the eigenvalues are in
this case necessarily distinct, one has that $M$ must be a  combination
of $A$ and $I$. Then, thanks to Lemma 4, the DS in NF takes just the form
of (20):
$$\.u=Au+\a(u)Au+\mu(u)u$$
If $\mu=0$, the NF satisfies Condition "A" and there is a convergent
normalizing \tr . If $\mu\ne 0$, the same result follows from
Theorem $1'$: indeed, there are no  \coms\ for the NF, and the only
admitted symmetry is $Au$; then the argument proceeds just as before.
\bs
It can be remarked that in the case of dimension $n=2$, one gets directly
$\ell=1$, and the occurrence of  the special form (20)
of the NF is guaranteed by Lemma 4. Then, all assumptions are
automatically fulfilled in dimension $n=2$.
Instead, if $n>2$ and the NF has the general form (19),
it is clear that a \com\ $\rho$ of the linear
part $\.u=Au$ may be a \com\ of the full problem $\.u=Au+\^F$, even if it
is not a \com\ of each one of the single problems $\.u=M_ju$.

One of the possibilities which can guarantee the special form (20) of the
NF, and   allow us to repeat the above argument
on the number of \coms , is the presence of some
additional symmetries $g_j=B_ju+G_j$: thanks to Lemma 2, one has that
$\^F$ must satisfy $\{B_ju,\^F\}=0$, and this condition may exclude some of
the matrices $M_j$ in the expression (19). The next
example will show this possibility, and will also be a good
illustration of the above discussion.
\pn
{\bf Example.}

Consider the space $R^{2m}$, and  put $u\equiv(x_1,\ldots, x_m,y_1,
\ldots,y_m)\in R^{2m}$;
assume that a Lie group $\Gamma$ acts "diagonally" on
both the $m$-dimensional spaces of the vectors $x$ and $y$
through the same linear representation $\D$:
i.e. $x\to x'=\D x$, $y\to y'=\D y$. Consider then a DS of
the following form
$$\.u= Au+F\eqno(22)$$
where
$$ A =\pmatrix{0 & I_m \cr
               -I_m & 0 } \eqno(22')$$
and assume that $F(u)$ admits the symmetries $B_iu$, where $B_i$ are the
(matrix representatives in the direct sum $\D\oplus\D$
of the) Lie generators of this group $\Gamma$ \big(the linear part $Au$
fulfils this symmetry requirement, so that the full DS
(22) admits this symmetry\big).
Let us assume here for concreteness (the general case could be relevant
for the study of Hamiltonian DS, see [17,18], but we do not consider here
this situation) the case $m=3$, $\Gamma=SO(3)$ and $\D$ its fundamental
representation, and consider the DS
$$\.u=Au+p(u)u\eqno(23)$$
where $p(u)$ is an analytic function depending on
the quantities, thanks to the $SO(3)$ symmetry, $x^2=(x,x)$,
$y^2=(y,y)$, $x\cdot y=(x,y)$ (the parentheses stand for the scalar
product in $R^3$). Once in NF, this DS takes the form
$$\.u=Au+\a Au+\mu u\eqno(24)$$
where $\a$ and $\mu$ are functions of $r^2=x^2+y^2$ only, thanks to
Proposition 1, which ensures that the linear symmetry $B_iu$ is
preserved, and to Lemma 4 as well. This NF has precisely the special
form of (20), and it is easy to check that there are no \coms\ for (24),
apart from $0$-degree rational functions as $(x_1^2+y_1^2)/(x_2^2+y_2^2)$.
It is important
to notice here that, if our problem would  not possess the symmetry
$SO(3)$, the NF (24) would contain many other matrices $M_i$
(according to Lemma 4), but that it is precisely the presence of the symmetry
$SO(3)$ which forces the NF to contain only  $A$ and the identity.
This seems to confirm the conjecture [9,10,17,18]
that the presence of a "sufficient" number of symmetries may be an essential
request in order to guarantee the convergence of a normalizing \tr .
Let us now assume
(this example can in fact be viewed as a multi-dimensional extension of an
example given in [10]) that in the original DS (22) the function $p$ is a
homogeneous polynomial of degree $k$ of the quantities $x^2,\ y^2,\
x\cdot y$: then the following vector function
$$g=r^{2k}u\eqno(25)$$
is a nontrivial analytic symmetry for the original DS (22) (in the case
that $p=(x^2+y^2)^k$, we choose e.g. $g=(x\cdot y)^k u$), indeed
$$\{Au+pu,r^{2k}Iu\}=\{Au,r^{2k}Iu\}+p(u\cdot\grad r^{2k}) u -
r^{2k} (u\cdot\grad p)u = 0 $$
and so we can conclude, e.g. from $i_2$) of Theorem $1''$ and Theorem $1'$,
that the DS can be normalized by a convergent \tr .

\titleb{Acknowledgment}

I am grateful to prof. A.D. Bruno for a very useful and clarifying
discussion, and for his kind interest in this argument. Prof. Bruno
informed me that he succeeded in extending, under suitable conditions,
the results in [10] to the case of dimension $n=3$. I am also indebted to
my friend Giuseppe Gaeta for carefully reading the manuscript and for
useful comments.

\vfill\eject

\Ref
[1] D. Arnal, M. Ben Ammar, and  G. Pinczon, Lett. Math.
Phys. {\bf 8}  (1984), 467-476

[2] V.I. Arnold, "Geometrical methods in the theory of
differential  equations", Springer, Berlin 1988

[3] V.I. Arnold, Yu.S. Il'yashenko, "Ordinary differential equations"; in
{\it Encyclopaedia of Mathematical Sciences - vol. I, Dynamical Systems I},
(D.V. Anosov and V.I. Arnold eds.), p. 1-148, Springer, Berlin 1988

[4] A. Bazzani, E. Todesco, G. Turchetti, G. Servizi, "A normal form
approach to the theory of nonlinear betatronic motion", CERN Report 94-02,
Geneva 1994

[5] H.W. Broer, "Formal normal
form theorems for vector fields and some consequences for bifurcations in
the volume preserving case", Berlin, Springer, 1981

[6] H.W. Broer, F. Takens, "Formally symmetric normal forms and
genericity", Dynamics Reported {\bf 2}  (1989), 39-59

[7] A.D. Bruno, "Local methods in nonlinear differential equations",
Springer, Berlin  1989

[8] A.D. Bruno, Divergence of the real normalizing \tr , Selecta Math.
(formerly Sovietica) {\bf 12}  (1993), 13-23

[9] A.D. Bruno, private communication

[10] A.D. Bruno and S. Walcher, Symmetries and convergence of
normalizing \tr s, J. Math. Anal. Appl. {\bf 183}  (1994), 571-576

[11] G. Cicogna, G. Gaeta: Lie point-symmetries and Poincar\'e normal forms
for dynamical systems, Journ. Phys. A: Math. Gen. {\bf 23}
(1990), L799-L802

[12] G. Cicogna, G. Gaeta: Poincar\'e normal forms and Lie point symmetries,
Journ. Phys. A: Math. Gen. {\bf 27}  (1994),  461-476

[13] G. Cicogna, G. Gaeta: Normal forms and nonlinear symmetries, Journ.
Phys. A: Math. Gen. {\bf 27} (1994), 7115-7124

[14] G. Cicogna, G.Gaeta, Approximate symmetries in dynamical
systems, Nuovo Cimento B, to be published

[15] R. Courant, D. Hilbert, "Methods of Mathematical Physics",
Interscience Publ., New York 1962

[16] C. Elphick, E. Tirapegui, M.E. Brachet, P. Coullet, and
G. Iooss, Simple global characterization for normal forms of singular
vector fields, Physica {\bf D  29}  (1987),  95-127

[17] H. Ito, Convergence of Birkhoff normal forms for integrable systems,
Comm. Math. Helv. {\bf 64} (1989), 412-461

[18] H. Ito, Integrability of Hamiltonian systems and Birkhoff normal forms
in the simple resonance case, Math. Ann. {\bf 292} (1992), 411-444

[19] L.M. Markhashov, On the reduction of differential equations to the
normal form by an analytic \tr , J. Appl. Math. Mech. {\bf 38}
(1974),  788-790

[20] P.J. Olver, "Applications of Lie groups to differential
equations", Springer, Berlin, 1986

[21] L.V. Ovsjannikov, "Group properties of differential equations",
Novosibirsk, 1962;(English transl. by G.W. Bluman , 1967); and
"Group analysis of differential equations", Academic
Press, New York, 1982

[22] J.-C. van der Meer, "The Hamiltonian Hopf bifurcation", Berlin,
Springer, 1985

[23] S. Walcher, On differential equations in normal form, Math. Ann.
{\bf 291} (1991), 293-314

\bye